\documentclass[11pt]{article}
\usepackage{latexsym}

\def\be{\begin{equation}}
\def\ee{\end{equation}}
\def\bea{\begin{eqnarray}}
\def\eea{\end{eqnarray}}

\def\case#1/#2{\textstyle\frac{#1}{#2}}

\begin{document}
\begin{titlepage}

\vspace{.7in}

\begin{center}
\Large
{Interacting vector fields in Relativity without Relativity}
\\
\vspace{.7in} \normalsize \large{Edward Anderson$^1$ and Julian
Barbour$^2$}
\\
\normalsize
\vspace{.4in}
{\em $^1$ Astronomy Unit, School of Mathematical Sciences, \\
Queen Mary, Mile End Road, London E1 4NS, U.K. }

{\em $^2$ College Farm, South Newington, Banbury, Oxon OX15 4JG, U.K.}
\vspace{.2in}
\end{center}
\vspace{.3in}
\baselineskip=24pt

\begin{abstract}
\noindent Barbour, Foster and \'{O} Murchadha have recently
developed a new framework, called here the {\it{ 3-space
approach}}, for the formulation of classical bosonic dynamics.
Neither time nor a locally Minkowskian structure of spacetime are
presupposed. Both arise as emergent features of the world from
geodesic-type dynamics on a space of 3-dimensional metric--matter
configurations. In fact gravity, the universal light cone and
Abelian gauge theory minimally coupled to gravity all arise
naturally through a single common mechanism. It yields relativity
-- and more -- without presupposing relativity. This paper
completes the recovery of the presently known bosonic sector
within the 3-space approach. We show, for a rather general ansatz,
that 3-vector fields can interact among themselves only as
Yang--Mills fields minimally coupled to gravity.

\end{abstract}

\vspace{.3in}
Electronic addresses: $^1$eda@maths.qmw.ac.uk, $^2$julian@platonia.com

\end{titlepage}

\section{Introduction}
This paper develops further the {\it{3-space approach}} to
relativity and gauge theory introduced by Barbour, Foster and
\'{O} Murchadha \cite{BOF1}. In Sec. $2$, we recall
briefly the principles of Dirac's generalized Hamiltonian dynamics
\cite{Dirac}, on which the 3-space approach is based. We then give
Arnowitt, Deser and Misner's (ADM) \cite{ADM} $(3 + 1)$
reformulation of $4$-dimensional general relativity (GR) as an
example. This recasts GR as geometrodynamics \cite{Wheeler}: the
dynamics of Riemannian $3$-geometries. The 3-space approach
proceeds in the opposite direction, recovering GR coupled to the
matter fields of nature from 3-dimensional dynamical principles
alone. Some work has been done on this by Hojman, Kucha\v{r} and
Teitelboim (HKT) \cite{HKT} and by Teitelboim \cite{Teitelboim}, 
but they presupposed that the dynamics unfolds in
spacetime. This is unnecessary.

To set the scene, Sec. 3 outlines the 3-space
approach, which consists largely of a systematic examination of
Baierlein--Sharp--Wheeler (BSW) \cite{BSW} geodesic-type
actions. We identify the two key principles of such actions:
{\it{best matching}} (a universal method to implement
{\it{3-dimensional}} diffeomorphism invariance) and a {\it{local}}
square root (taken at each space point before integration over
3-space). These two principles replace Einstein's assumptions of
{\it{4-dimensional}} diffeomorphism invariance (general
covariance) and a locally Minkowskian structure of spacetime. 
The 3-space principles have the following consequences.

First, they essentially single out pure GR as the theory that
arises from considering 3-geometries alone. Second, if a scalar
field on the 3-geometries is included, they force it to obey the
same light-cone structure as gravity. Both minimal and
Brans--Dicke couplings are allowed \cite{BOF1}.
Third, and even more
remarkably, if a 3-vector field is included, it must not only
respect the gravitational light-cone but also be massless and
satisfy the Maxwell equations of electromagnetism
minimally-coupled to gravity. Finally they show in a forthcoming paper that such a 3-vector field can
couple to scalar fields only through Abelian U(1) gauge theory.
The new approach leads to these sharp predictions because it uses
a spare ontology (3-space replaces spacetime) and employs Dirac's
powerful generalized Hamiltonian dynamics \cite{Dirac} to
construct geodesic-type actions on the arena of configuration
space, not spacetime. We note that quantum mechanics also unfolds
on configuration space.

In Sec. $4$, we extend the existing 3-space results to 3-vector
fields allowed to interact amongst themselves. We find that
Yang--Mills theory \cite{YM} minimally coupled to GR is the only
possibility allowed for quite a general ansatz for the 3-vector
fields' potential. More precisely, the 3-vector fields must again
respect the gravitational light-cone, be fundamentally massless,
and have a Yang--Mills type mutual interaction. Our current formalism
can not predict how many vector fields there are in nature, nor what 
their gauge groups are.

In Sec. 5 we conclude that, within the bosonic sector, the 3-space
approach yields the key features of the observed world. Gravity,
the universal light cone, and gauge theory all arise in
essentially the same manner through the single mechanism of
consistent Dirac-type constraint propagation applied to the
interplay of best matching with the local square root. 
We finally consider
whether classical topological terms can be accommodated in our
formalism. These would usually play a part in the interpretation
of quantum chromodynamics (QCD) \cite{Hooft,Peccei,PS,Weinberg}.

\section{Hamiltonian Dynamics and General Relativity}

Let $\mbox{\sffamily L} \normalfont (q_{\Theta},
\dot{q}_{\Theta})$ be a Lagrangian provisionally adopted for some
general theory of tensor fields $q_{\Theta}(x,\lambda)$ \footnote{
In this paper, $x$ is a 3-dimensional spatial argument, 
$\lambda$ is a time label and
$\frac{\partial}{\partial\lambda}$ is denoted by a dot.  
We use capital Greek letters as indexing sets for the fields;
the number of fields in the set indexed by $\Theta$ is denoted by $\#\Theta$.
We use lower-case Greek letters for spacetime indices and lower-case Latin letters
for spatial indices. 
We use capital Latin letters for internal indices; 
no significance is attached to whether these are raised or lowered
but their order will be important.
We use round brackets to denote symmetrization of indices,
and square brackets to denote antisymmetrization.  
Indices unaffected by the (anti)symmetrization are set between vertical lines
}.  If not all the conjugate momenta $p^{\Theta} = \frac
{\partial \mbox{\sffamily \scriptsize L \normalsize\normalfont}}  {\partial
\dot{q}_{\Theta}}  \normalfont$ can be inverted to
give the $\dot{q}_{\Theta}$ in terms of the $p^{\Theta}$, then the
theory has primary constraints ${\cal C }_{\Pi}(q_{\Theta},
p^{\Theta}) = 0$ solely by virtue of the form of \mbox{\sffamily L
\normalfont}.  As Dirac noted \cite{Dirac}, in such a case a
theory described by a Hamiltonian $\mbox{\sffamily
H\normalfont}(q_{\Theta}, p^{\Theta})$ could just as well be
described by a Hamiltonian \be \mbox{\sffamily
H\normalfont}_{\mbox{\scriptsize Total  \normalsize}} =
\mbox{\sffamily H\normalfont} + N_{\Pi} {\cal C } ^ {\Pi}
\label{hamtot} \ee for arbitrary functions $N_{\Pi}$.  Moreover,
one needs to check that the primary constraints and any further
secondary constraints ${\cal C }_{\Gamma}(q_{\Theta}, p^{\Theta})$
(obtained as true variational equations ${\cal C}_{\Gamma} = 0$)
are propagated by the evolution equations.  If they are, then the
constraint algebra indexed by $\Delta_{(1)} = \Pi \bigcup \Gamma$
closes, and a classically-consistent theory is obtained.  This
happens when $\dot{{\cal C }}_{\Delta_{(1)}}$ vanishes either due to 
the Euler--Lagrange equations alone 
or additionally due to the vanishing of ${\cal C }_{\Delta_{(1)}}$, 
which is Dirac's notion of weak vanishing, 
denoted by $\dot{{\cal C}}_{\Delta_{(1)}} \approx 0$.

If ${\cal C }_{\Delta_{(1)}}$ does not vanish weakly, then it must
contain further independently-vanishing expressions ${\cal C
}_{\Sigma_{(1)}}(q_{\Theta}, p^{\Theta})$ in order for the theory
to be consistent.  One must then enlarge the indexing set to
$\Delta_{(2)} = \Delta_{(1)} \bigcup \Sigma_{(1)}$ and see if
$\dot{{\cal C }}_{\Delta_{(2)}} \approx 0$.  In principle, this
becomes an iterative process by which one may construct a full
constraint algebra indexed by $\Delta_{\mbox {\scriptsize (final)
\normalsize}}$ by successive enlargements 

\noindent $\Delta_{(i+1)} = \Delta_{(i)} \bigcup \Sigma_{(i)}$. In practice,
however, the process cannot continue for many steps since

\noindent $\#\Delta_{(i + 1)} > \#\Delta_{(i)}$, $\#\Theta$ is a small
number, and we need the true number of degrees of freedom to
satisfy $\#\Theta - \#\Delta_{\mbox{\scriptsize{(final)}
\normalsize}}
> 0$ to have any nontrivial theory.  It should be emphasized that
there is no guarantee that a given Lagrangian will give rise to
any consistent theory.

In the case of $4$-dimensional GR, \mbox{\sffamily H\normalfont}
is zero, but the $(3 + 1)$ ADM split \cite{ADM} yields the
gravitational case of (2), $ \mbox{\sffamily
H\normalfont}_{\mbox{\scriptsize Total \normalsize}} = N{\cal H }
+ N^i {\cal H }_i \approx 0$, where ${\cal H}$ is the Hamiltonian
constraint, ${\cal H}_i$ is the momentum constraint and the arbitrary
functions $N$ and $N^i$ are the lapse and the shift. The Bianchi
identities then ensure that $\dot{{\cal H}_i} \approx 0 $ and
$\dot{{\cal H}} \approx 0 $ so that ${\cal H}$ and ${\cal H}_i$
form a closed constraint algebra: the Dirac algebra.  The first
propagation corresponds to the invariance of the ADM action under
$3$-diffeomorphisms.  The second propagation corresponds to a
remarkable hidden symmetry of GR, foliation invariance, which is
the invariance under local reparametrization of the time label.
The $(3 + 1)$ split can also be done in the presence of matter
fields \cite{ADM, Kuchar} indexed by $\Psi$. We will denote the
Hamiltonian and momentum constraints obtained in this case by
$^{\Psi}{\cal H}$ and $^{\Psi}{\cal H } _i$.

Whereas ADM decomposed $4$-dimensional spacetime in a (3 + 1)
split, the work of HKT \cite{HKT} and of Teitelboim
\cite{Teitelboim} goes in the opposite direction.  They
reconstruct $4$-dimensional GR with matter fields from
geometrodynamics by requiring that the constraint algebra of
$^{\Psi}{\cal H}$ and $^{\Psi}{\cal H } _i$ closes to reproduce
the Dirac algebra.

\section{Relativity without Relativity}

Barbour, Foster and \'{O} Murchadha
(BF\'{O})\cite{BOF1} have recently used Dirac's
general procedure under the much weaker assumption that the
constraint algebra merely closes.  The requirement that the Dirac
algebra be reproduced imports the foliability of $4$-dimensional
spacetime. BF\'{O} showed that this is largely unnecessary if one
wishes to derive GR from $3$-dimensional principles.  They use
truly $3$-dimensional principles alone, so we can call it the
3-space approach. It gives new insights into the origin of both
special and general relativity, and furthermore, of Abelian gauge
theory.

Their point of departure was a paper of Baierlein, Sharp and
Wheeler \cite{BSW}, in which it is shown that the
Einstein--Hilbert action $\int d^4x \sqrt{-g^{(4)}}R^{(4)}$ can,
in the case of globally-hyperbolic spacetimes, be rewritten in the
three-dimensional form \be S_{\mbox{\scriptsize BSW \normalsize}}
= \int d\lambda \int d^3x \sqrt{g} \sqrt{R}
\sqrt{T_{\mbox{\scriptsize g\normalsize}}}, \label{ABSW} \ee where
$g$ is the determinant of the spatial $3$-metric $g_{ij}$ induced
on spacelike hypersurfaces by the $4$-metric $g^{(4)}_{\mu\nu}$, 
and $R$ is the $3$-dimensional  Ricci scalar formed from $g_{ij}$. In
principle this action defines a measure on the quotient space
\be 
\mbox{\{Superspace\}} = \frac {\mbox {\{Riem\}} }
{\mbox {\{Diffeomorphisms\}} }, 
\ee 
where $\mbox{\{Riem\}}$ is the
space of all Riemannian $3$-geometries on some given topology; we
will work throughout this paper with compact topologies without boundary. 
The label $\lambda$ parametrizes some chosen curve of 3-metrics, which is a
primary object in superspace. The gravitational kinetic term
$T_{\mbox{\scriptsize g\normalsize}}$ is given by \be
T_{\mbox{\scriptsize g\normalsize}} = G^{abcd}(\dot{g}_{ab} -
\nabla_a\xi_b - \nabla_b\xi_a) ( \dot{g}_{cd} - \nabla_c\xi_d -
\nabla_d\xi_c), \ee where $G^{abcd} = g^{ac}g^{bd} - g^{ab}g^{cd}$
is the DeWitt supermetric \cite{DeWitt}, $\xi_a = g_{ab}N^b$
\footnote{ $\xi_i$ is formally a velocity, so
$T_{\mbox{\scriptsize g\normalsize}}$ is homogeneously quadratic
in its velocities. The interpretation of $\xi_i$ and its
importance in gauge theory are explained in \cite{Barbour}.} and
$\nabla_a$ (or $_{;a}$) denotes the covariant derivative with
respect to $g_{ij}$ .

To compensate for the possible coordinate change in going between
neighbouring $3$-geometries, correction terms have been added to
each of the bare velocities $\dot{g}_{ab}$ in $T_g$. One can think
of this as Bertotti and Barbour's method for achieving
$3$-diffeomorphism invariance \cite{BB,BOF1}. It is appropriately
called {\it{best matching}} because the variation with respect to
$\xi_i$ can be seen as implementing a best matching of two
infinitesimally-differing $3$-dimensional configurations on a
compact manifold; the aim of this is to bring the two
configurations as close as possible to congruence and then define
the residual difference between them as a measure on superspace.
Three-diffeomorphism invariance is achieved by the prescription,
valid and uniquely defined for any bosonic field $B$, \be
\mbox{bare velocity } \dot{B} \longrightarrow \mbox{best-matched
velocity } \dot{B} - \pounds_{\xi}B, \ee where $\pounds_{\xi}$ is
the Lie derivative with respect to the
$3$-diffeomorphism-generating auxiliary field $\xi_i$.
Three-diffeomorphism invariance is an example of a gauge symmetry
leading to a constraint that is homogeneously linear in the
momenta.  Whenever a theory has this form of gauge symmetry, some
corresponding form of best matching occurs. Because
3-diffeomorphisms must, in a Machian framework \cite{BB}, be
applied to all fields, both metric and material, the best matching
that implements them has far-reaching universal consequences, as
we shall see.

The BSW action resembles Jacobi's action principle \cite{BOF1}
for the (timeless) dynamical orbit of a Newtonian $N$-body system
in its $3N$-dimensional configuration space \cite{Lanczos}, 
but differs from it in that the latter contains a single square root, 
whereas the former has one square root at each space point, 
after which these are integrated over
all space.  
We call the latter choice the \it local \normalfont square root.
The presence of the square root means that the
Lagrangian is homogeneous of degree $1$ in the velocities, so that
the canonical momenta must be homogeneous of degree $0$. As Dirac
noted \cite{Dirac}, such canonical momenta must satisfy at least
one primary constraint as an identity. 

For the BSW Lagrangian, the canonical
momenta (defined at each space point) are 
\be p^{ij} = \frac{\partial\mbox{\sffamily{L}\normalfont} }       
{ \partial\dot{g}_{ij}} 
= \sqrt{\frac{gR}{T}}(g^{ic}g^{jd} - g^{ij}g^{cd})(\dot{g}_{cd} - \nabla_c\xi_d - \nabla_d\xi_c)
\ee 
and the primary constraint that holds at each space point is
\be 
g{\cal H } \equiv gR - p^{ij}p_{ij} + \frac{1}{2}p^2 = 0
\label{GRHam}, 
\ee 
where $p$ denotes the trace of $p^{ij}$.  
In addition, variation of the BSW action with respect to $\xi_i$
leads to the secondary momentum constraint 
\be 
\frac{1}{2}\sqrt{g}{\cal H}^j \equiv {p^{ij}}_{;i} = 0 \label{GRMom}. 
\ee 
(\ref{GRHam}) and (\ref{GRMom}) are respectively 
the Hamiltonian and momentum constraints of GR.  
The corresponding Euler--Lagrange equations ensure that these propagate, 
so the constraint algebra is closed. 
At first glance, one would expect the BSW action to be invariant 
only with respect to the global reparametrization 
$\lambda \longrightarrow \lambda^{\prime}(\lambda)$, 
for $\lambda^{\prime}$ a monotonic arbitrary function of $\lambda$.  
But in fact the action is invariant under the far more general local transformation 
\be
\lambda \longrightarrow \lambda^{\prime}(\lambda),\mbox{  }
g_{ij}(x,\lambda) \longrightarrow g_{ij}(x,\lambda^{\prime}),
\mbox{ } \xi_i(x,\lambda) \longrightarrow
\frac{d\lambda^{\prime}}{d\lambda}\xi_i(x,\lambda). 
\ee

This remarkable invariance does not hold for the generalization of
the BSW action that BF\'{O} started with: \be S_{\mbox{\scriptsize BFO
\normalsize}} = \int d\lambda \int d^3x \sqrt{g} \sqrt{\Lambda +
P(x,\lambda)} \sqrt{T_{\mbox{\scriptsize W\normalsize}}}
\label{BFOac}, \ee where $\Lambda$ is an arbitrary constant, the
potential $P$ is some arbitrary scalar formed from $g_{ij}$ and
its spatial derivatives up to a given order, and
$T_{\mbox{\scriptsize W\normalsize}}$ is the same as
$T_{\mbox{\scriptsize g\normalsize}}$ except that it contains a
generalized supermetric, $G_{\mbox{\scriptsize
W\normalsize}}^{abcd} = g^{ac}g^{bd} - Wg^{ab}g^{cd}$.  Their
first result is as follows. The action (\ref{BFOac}) is defined
solely in terms of $3$-dimensional concepts, and associates an
action with curves on the space $\mbox{\{Riem\}} \times \Xi$,
where $\Xi$ is the vector space to which $\xi_i$ belongs.  Then,
the presence of the local square root in (\ref{BFOac}) gives the
primary constraint \be g{\cal H } \equiv g(P + \Lambda) -
p^{ij}p_{ij} + \frac{2W}{2(3W - 1)}p^2 = 0 \label{BFOHam} \ee and
variation with respect to $\xi_i$ leads to an unchanged secondary
constraint, (\ref{GRMom}). The latter can be regarded as a
differential equation for $\xi_i$ (which is contained in
$p^{ij}$). If this can be solved (for the issues, as yet not fully
resolved, that are involved, see the papers \cite{thin sandwich}),
the action will depend only on the curve in superspace. This
follows from the constraints being free of $\xi_i$, and the
momentum constraint reducing the number of degrees of freedom to
$3$, which is the number of degrees of freedom per space point in
a $3$-geometry.

The question posed in \cite{BOF1} -- and answered in the
affirmative -- is whether GR can be derived solely from
$3$-dimensional arguments, that is, without any recourse to
arguments related to $4$-dimensional general covariance.  The
approach succeeds because of the need to propagate the quadratic
constraint ${\cal H }$ acts as a powerful filter of viable
theories, which are already strongly restricted by the universal
linear 3-diffeomorphism constraint \footnote{Throughout this
paper, the momentum constraints are automatically propagated
because the action has been deliberately constructed to be
invariant under ($\lambda$-dependent) $3$-diffeomorphisms.}.  Up
to second-order spatial derivatives of $g_{ij}$, this works for 
\be 
S_{\mbox{\scriptsize BFO \normalsize}} = 
\int d\lambda\int d^3x \sqrt{g} \sqrt{\Lambda + \epsilon R}
\sqrt{T_{(\mbox{\scriptsize W\normalsize} = 1)}}
\label{BFOallowed}, 
\ee 
where $\epsilon \in \{-1, 0, 1 \}$, and the
subscript of $T$ indicates that the a priori free parameter $W$
must take the DeWitt value, $1$. The cases $\epsilon = 1$ and
$\epsilon = -1$ correspond to Lorentzian and Euclidean GR
respectively. The case $\epsilon = 0$ is called {\it{strong
gravity}} \cite{strong}, because it is the limit as Newton's
gravitational constant, $\kappa$, goes to infinity of the other
two cases. This is a theory which is 4-dimensionally generally
covariant in the sense of having four constraints per 3-space
point but cannot, unlike the other two cases, be represented by
tensorial equations on a 4-dimensional spacetime manifold. This
signature freedom $\epsilon$ and the freedom to have a $\Lambda$
(which we identify as a cosmological constant) is what we mean in
the introduction by GR being essentially singled out. Furthermore,
all the higher-derivative corrections considered in \cite{BOF1}
were found not to give a propagating ${\cal H }$.  However, a conformal
generalization of the above work (which has not yet be fully
elaborated) is also possible, giving another theory which has no
spacetime manifold interpretation \cite{conformal}. In fact, we
anticipate that the full significance of the 3-space approach will
not become apparent until the full generalizations to conformal
superspace, which will result in a fully scale-invariant theory,
and the fermionic sector have been completed.

Barbour, Foster and \'{O} Murchadha included a scalar field $\phi$
by considering the action \be S_{\mbox{\scriptsize
BSW\normalsize}_{\phi}} = \int d\lambda \int d^3x \sqrt{g} \sqrt{
R + U_{\phi}} \sqrt{T_{\mbox{\scriptsize g\normalsize}} +
T_{\phi}} \ee with the gravitationally best-matched scalar kinetic
term $T_{\phi} = (\dot{\phi} - \pounds_{\xi}\phi)^2$ and the
potential ansatz $U_{\phi} = -(C/4)g^{ab}\phi_{,a}\phi_{,b} +
\sum_{(n)} A_{(n)}\phi^n$. Then the square local root gives as an
identity the primary constraint \be g ^{\phi}{\cal H } \equiv g(R
+ U_{\phi}) - p^{ij}p_{ij} + \frac{1}{2}p^2 - \pi^2 = 0, \ee where
$\pi$ is the momentum conjugate to $\phi$.  Variation with respect
to ${\xi}_i$ gives the momentum constraint \be \frac{1}{2}\sqrt{g}
^{\phi}{\cal H}^i = {p^{ij}}_{;j} - \frac{1}{2} \pi\phi^{,i} = 0.
\ee The constraint $^{\phi}{\cal H }$ contains the canonical
propagation speed $C$ of the scalar field.  A priori, $ C \neq 1
$, which means there is no reason for the scalar field to obey the
same light-cone as gravity. However, imposing $^{\phi}\dot{{\cal H
} } \approx 0$ gives a putative secondary constraint \be \frac{(1
- C)}{N}(N^2\pi\phi_{;i})^{;i} = 0 \ee and the theory has just one
scalar degree of freedom, so if the cofactor of $(1 - C)$ were
zero, the scalar dynamics would be trivial. Thus one has derived
that $C = 1$: scalar fields must obey the gravitational
light-cone. Notice also that this scheme gives minimal coupling of
the scalar field to gravity. However, there is one other
possibility because the most general kinetic term includes also a
metric--scalar cross-term. This gives Brans--Dicke theory \cite{BOF1}.

For our use in Sec. 4, we now sharpen up the procedure used by BF\'{O} 
on inclusion of a single vector field.
They considered the action
\be 
S_{{ \mbox{\scriptsize BSW\normalsize}_{\mbox{\tiny
A\normalsize}}}} = \int d\lambda \int d^3x \sqrt{g} 
\sqrt{R + U_{\mbox{\scriptsize A\normalsize}}} 
\sqrt{T_{\mbox{\scriptsize g\normalsize}} + T_{\mbox{\scriptsize A\normalsize}}},  
\ee 
for
$T_{\mbox{\scriptsize A\normalsize}} 
= g^{ab}(\dot{A}_a - \pounds_{\xi}A_a)(\dot{A}_b - \pounds_{\xi}A_b)$ 
the quadratic gravitationally best-matched kinetic term of $A_a$,\footnote{The
Lie derivative of $A_a$ with respect to $\xi_i$ is
$\pounds_{\xi}A_a =  {A_a}^{;c}\xi_c + A_c {\xi^c}_{;a}$ .} and
the potential ansatz $U_{\mbox{\scriptsize A\normalsize}} =
C_1A_{a;b}A^{a;b} + C_2A_{a;b}A^{b;a} + C_3{A_a}^{;a}{A_b}^{;b} +
\sum_{(k)} B_{(k)}(A_aA^a)^{k} $. Then, the local square root gives as
an identity the primary constraint \be g ^{\mbox{\scriptsize
A\normalsize}}{\cal H } \equiv g(R + U_{\mbox{\scriptsize
A\normalsize}}) - p^{ij}p_{ij} + \frac{1}{2}p^2 - \pi^a\pi_a = 0,
\ee where $\pi^a$ is the momentum conjugate to $A_a$. Variation
with respect to $\xi_i$ gives the momentum constraint \be
\frac{1}{2}\sqrt{g} ^{\mbox{\scriptsize A\normalsize}}{\cal H}^j =
{p^{ij}}_{;i} - \frac{1}{2}(\pi^{c}({A_{c}}^{;j} - {{A}^j}_{;c}) -
{\pi^{c}}_{;c}A^{j}) = 0. \ee Then, imposition of
$^{\mbox{\scriptsize A\normalsize}}\dot{{\cal H }} \approx 0$
gives rise to \be
 \frac{4C_1 + 1}{N} (N^2\pi^a{A_a}^{;b})_{;b}
+\frac{4C_2 - 1}{N} (N^2\pi^a{A^b}_{;a})_{;b} +\frac{4C_3}{N}
(N^2\pi^b{A^a}_{;a})_{;b} -\frac{1}{N} (N^2{\pi^a}_{;a}A^b)_{;b} =
0. \ee This gives a nontrivial theory only for $C_1 = -C_2 = -1/4$,
$C_3 = 0$ and if there is a secondary constraint ${\cal G} =
{\pi^a}_{;a}$. The conditions on the $C$'s mean that the
gravitational light-cone is obeyed, and furthermore that the
derivative terms in $U_{\mbox{\scriptsize A\normalsize}}$ are
$-(curl\mbox{\bf{A} \normalfont})^2/4$.  One then requires that
$\dot{{\cal G }} \approx 0$, which gives rise to $
2\sqrt{g}(N\Sigma_{(k)}B_{(k)}(A_aA^a)^{(k - 1)}A^i)_{;i} = 0$.
This forces the $B_{(k)}$ to be zero. In particular, $B_{(1)} = 0$
means that the vector field must be massless.  The working for
these two steps is a subcase of that in Sec. $4$, so the
details of the calculation are implicitly contained there.

Now, the form of $U_{\mbox{\scriptsize A\normalsize}}$ allowed is
invariant under the gauge transformation $ A_a \longrightarrow A_a
+ \nabla_a\Lambda $, so we are dealing with a gauge theory. Thus,
if we introduce an auxiliary variable $\Phi$ into
$T_{\mbox{\scriptsize A\normalsize}}$ such that variation with
respect to it encodes ${\cal G }$, then we should do so according
to the best matching corresponding to this gauge symmetry. This
uniquely fixes the form of $T_{\mbox{\scriptsize A\normalsize}}(A,
\Phi)$ to be \be T_{\mbox{\scriptsize A\normalsize}} = (\dot{A}_a
- \pounds_{\xi}A_a - \Phi_{,a})(\dot{A}^a - \pounds_{\xi}A^a -
\Phi^{,a}). \ee Thus, if one identifies $\Phi$ as $A_0$, this
derivation forces $A_{\alpha} = [A_0, A_a]$ 
to obey Maxwell's equations minimally-coupled
to gravity \cite{Kuchar}. Moreover, as noted by Giulini
\cite{giulini} and reported in \cite{BOF1}, the massive vector
field does not fit in the conceptual scheme of the 3-space
approach although it is a generally covariant theory. We see that
the 3-space approach does not yield all generally covariant
theories. But \cite{BOF1} and the present paper show that it does
at least yield the bosonic fields hitherto observed in nature.

Finally, on attempting to couple $A_a$ to scalar fields by the
inclusion of interaction terms, BF\'{O} have shown similarly that
demanding the propagation of $ ^{\mbox{\scriptsize
A\normalsize}\phi}{\cal H}$, and of any secondary constraints
arising from it, leads to $U(1)$ gauge theory. We have thus a
chain of successively more sophisticated theories, each arising
from its predecessor by iteration of constraint propagation
consistency. This not only prises open the door to classical
physics.  It shows one a way to derive it. Thus, Dirac's work,
applied to best matching theories with local square roots in their
actions, leads to a striking alternative to Einstein and
Minkowski's $4$-dimensional foundation of physics. The above
outline of `Relativity without Relativity' will make the remainder
of this paper into an almost algorithmic formality, from which
Yang--Mills gauge theory will emerge from allowing a general
collection of $3$-vector fields to interact with each other.

\section{$K$ Interacting Vector Fields}

We consider a BSW-type action containing the a priori
unrestricted vector fields $A_a^I$ ($I = 1$ to $K$), 
\be S_{\mbox{\scriptsize BSW\normalsize}_{{\mbox{\tiny A\normalsize}}_I}} = \int d\lambda
\int d^3x \sqrt{g} \mbox{{\sffamily L\normalfont}} (g_{ij},
\dot{g}^{ij}, A_i^I, \dot{A}^i_I, N, N^i) = \int d\lambda \int
d^3x \sqrt{g} \sqrt{R + U_{\mbox{\scriptsize A\normalsize}_I}}
\sqrt{ T_{\mbox{\scriptsize g\normalsize}} + T_{\mbox{\scriptsize
A\normalsize}_I}}. \ee 
We use the most general homogeneous quadratic best-matched kinetic term
$T_{\mbox{\scriptsize A\normalsize}_I}$, and a general ansatz for
the potential term $U_{\mbox{\scriptsize A\normalsize}_I}$. We
could have constructed these using the inverse $3$-metric $g^{ab}$
as the only possible means of contracting spatial indices. But,
for greater generality, we have also introduced the totally
antisymmetric tensor density $\epsilon^{abc}$ for this purpose.

We note that no kinetic cross-term $\dot{g}_{ab}\dot{A}_{Ic}$ is
possible. This is because the only way to contract 3 spatial
indices is to use $\epsilon^{abc}$, and $\dot{g}_{ab}$ is
symmetric. Then $T_{\mbox{\scriptsize A\normalsize}_I}$ is
unambiguously \be T_{\mbox{\scriptsize A\normalsize}_I} =
P_{IJ}g^{ad}( \dot{A}_a^I - \pounds_{\xi}A_a^I)( \dot{A}^J_d -
\pounds_{\xi}A^J_d), 
\ee 
for $P_{IJ}$ without loss of generality a symmetric constant matrix.  
We will assume that $P_{IJ}$ is positive-definite 
so that the quantum theory of $A^I_a$ has a well-behaved inner product.  
In this case, we can take $ P_{IJ} = \delta_{IJ} $ by rescaling the vector fields.

We consider the most general $U_{\mbox{\scriptsize A\normalsize}_I}$ 
up to first derivatives of $A_{Ia}$, and up to four spatial index contractions.  
We note that the latter is equivalent to the necessary naive power-counting requirement for
the renormalizability of any emergent four-dimensional quantum
field theory for $A_{Ia}$.  
Then $U_{\mbox{\scriptsize A\normalsize}_I}$ has the form 
\be
\begin{array}{ll}
U_{\mbox{\scriptsize A\normalsize}_I} = & O_{IK}C
^{abcd}A^{I}_{a;b}A^K_{c;d} + {B^I}_{JK}\bar{C
}^{abcd}A_{Ia;b}A^J_c A^K_d + I_{JKLM}\bar{\bar{C
}}^{abcd}A^J_aA^K_bA^L_cA^M_d \\ & +
\frac{1}{\sqrt{g}}\epsilon^{abc}(D_{IK}A^I_{a;b}A^K_c +
E_{IJK}A^I_aA^J_bA^K_c) + F_Ig^{ab}A^I_{a;b} +
M_{IK}g^{ab}A^I_aA^K_b,
\end{array}
\ee where $C^{abcd} = C_1g^{ac}g^{bd} + C_2g^{ad}g^{bc} +
C_3g^{ab}g^{cd}$ is a generalized supermetric, and similarly for
$\bar{C}$ and $\bar{\bar{C}}$ with distinct coefficients. 
$O_{IK}$, $B_{IJK}$, $I_{JKLM}$, $D_{IK}$, $E_{IJK}$, $F_{I}$, $M_{IK}$
are constant arbitrary arrays.  
Without loss of generality $O_{IK}$, $M_{IK}$ are symmetric 
and $E_{IJK}$ \normalfont is totally antisymmetric.

Defining $2N = \sqrt {(T_{\mbox{\scriptsize g\normalsize}} +
T_{\mbox{\scriptsize A\normalsize}_I})/ (R + U_{\mbox{\scriptsize
A\normalsize}_I})}$, the conjugate momenta are given by 
\be p^{ij} = \frac{\partial\mbox{\sffamily{L}\normalfont} }       { \partial
\dot{g}_{ij}} = \frac{\sqrt{g}}{2N}(g^{ic}g^{jd} -
g^{ij}g^{cd})(\dot{g}_{cd} - \nabla_c\xi_d - \nabla_d\xi_c), 
\ee
\be 
\pi_I^i = \frac   {\partial\mbox{\sffamily{L}\normalfont}}
{\partial \dot{A^I_i}}  = \frac{\sqrt{g}}{2N} ( \dot{A_I^i}
-\pounds_{\xi}A_I^i), 
\ee 
which can be inverted to give expressions
for $\dot{g}_{ij}$ and $\dot{A}^I_i$. The local square root gives
as an identity the primary Hamiltonian constraint, \be g
^{\mbox{\scriptsize A\normalsize}_I}{\cal H } = g(R +
U_{\mbox{\scriptsize A\normalsize}_I}) - p^{ij}p_{ij} +
\frac{1}{2}p^2 -\pi^I_i\pi_I^i   = 0. \ee
  We get the secondary momentum constraint by varying with respect to $\xi$:
\be \frac{1}{2}\sqrt{g} ^{\mbox{\scriptsize A\normalsize}_I}{\cal
H}^j = {p^{ij}}_{;i} - \frac{1}{2}(\pi^{Ic}({A_{Ic}}^{;j} -
{{A_I}^j}_{;c}) - {\pi_I^c}_{;c}A^{Ij}) = 0. \ee

The Euler--Lagrange equations for $g_{ij}$ and $A^I_i$ are 
\be
\begin{array}{ll}
\frac{\partial p^{ij}} {\partial \lambda} =
\frac{\delta\mbox{\sffamily\scriptsize
L\normalfont\normalsize}}{\delta g_{ij}} = & \sqrt{g}N(g^{ij}R -
R^{ij}) - \frac{2N}{\sqrt{g}}\left(p^{im}{p_m}^j -\frac{1}{2}p^{ij}p \right) +
\sqrt{g}(Ng^{ij}U_{\mbox{\scriptsize A\normalsize}_I} + N^{;ij}
-g^{ij}{\nabla}^2 N) \\ & - \frac{N}{\sqrt{g}}\pi^{Ii}{\pi_I}^j +
\pounds_{\xi}p^{ij} +
\sqrt{g}O^{IK}(N(2A_{I(b;d)}{A_K}^{(j|}C^{bde|i)} -
A_I^eA_{Kb;d}C^{bd(ij)}))_{;e} \\ & -\sqrt{g}NO^{IK}\left(
\begin{array}{l} C_1({A_I^i}_{;a} A_K^{j;a} + A_I^{a;i}
{A_{Ka}}^{;j})
+ C_2(A_I^{i;a} {A_{Ka}}^{;j} + {A_{Ia}}^{;j} A_K^{i;a}) \\
+ 2C_3 {A_I}^{(i;j)}{A_K^{a}}_{;a} \end{array} \right) \\ &
+ {B^I}_{JK} \frac{\sqrt{g}}{2}
\left( \begin{array}{l}(\bar{C_1} + \bar{C_2}) (N( {A_I}^{(i|}A^{J|j)}A^{Kb} +
{A_I}^{(i|}A^{Jb}A^{K|j)} - A_I^bA^{J(i|}A^{K|j)}))_{;b} \\
+ \bar{C_3}(2(N{A_I}^{(i|}A_c^JA^{Kc})^{;|j)} - g^{ij}(NA_I^bA^J_cA^{Kc})_{;b})\\
- 2N\bar{C}_1({A_I}^{(i|;b}A^{J|j)}A^K_b + {A_I}^{b;(i|}A^J_bA^{K|j)}) \\
- 2N\bar{C}_2({A_I}^{(i|;b}A^J_bA^{K|j)} + {A_I}^{b;(j|}A^{J|i)}A^K_b) \\
- 2N\bar{C}_3({A_I}^{(i;j)}A^{Jb}A^K_b + {A_I^b}_{;b}A^{J(i|}A^{K|j)}) \end{array} \right) \\ &
- N\sqrt{g}I_{JKLM} \left( \begin{array}{c}\bar{\bar{C}}_1(A^{J(i|}A^{Kb}A^{L|j)}A^M_b  
+ A^{Jb}A^{K(i|}A^L_bA^{M|j)})\\
+ \bar{\bar{C}}_2(A^{J(i|}A^{Kb}A^L_bA^{M|j)} + A^{Jb}A^{K(i|}A^{L|j)}A^M_b)  \\
+ \bar{\bar{C}}_3(A^{J(i|}A^{K|j)}A^{Lb}A^M_b +
A^{Jb}A^K_bA^{L(i|}A^{M|j)} ) \end{array} \right) \\ & -
\sqrt{g}F^I\left(N{A_I}^{(i;j)} - (N{A_I}^{(i})^{;j)}  +
\frac{1}{2}g^{ij}(NA_I^b)_{;b} \right) \\ & -
N\sqrt{g}M^{JK}{A_J}^{(i}{A_K}^{j)} - \frac {N}{2}
g^{ij}\epsilon^{abc}(D_{IK}A^I_{a;b} + E_{IJK}A^I_aA^J_b)A^K_c,
\end{array}
\ee

\be
\begin{array}{ll}
\frac{\partial \pi^{Ji}} {\partial \lambda} =
\frac{\delta\mbox{\sffamily\scriptsize
L\normalfont\normalsize}}{\delta A_{Ji}} = &
-2\sqrt{g}O^{JK}(C_1(N{A_K^i}_{;b})^{;b} +
C_2(N{A_{Kb}}^{;i})^{;b} +C_3(N{A_{Kb}}^{;b})^{;i}) \\ & +
\sqrt{g}( NA^I_{a;b}A_{Mc}(\bar{C}^{abci}{B_I}^{MJ} +
\bar{C}^{abic}{B_I}^{JM}) -
(NA^M_cA^K_d)_{;b}\bar{C}^{ibcd}{B^J}_{MK}) \\ & +
\sqrt{g}N(\bar{\bar{C}}^{ibcd}I^{JKLM} +
\bar{\bar{C}}^{bicd}I^{KJLM} + \bar{\bar{C}}^{bcid}I^{KLJM} +
\bar{\bar{C}}^{bcdi}I^{KLMJ})A_{Kb}A_{Lc}A_{Md} \\ & +
\epsilon^{abi}(D^{IJ}NA_{Ia;b} +D^{JI}(NA_{aI})_{;b}) +
3\epsilon^{ibc}E^{JNK}NA_{Kc}A_{Nb} \\ & - \sqrt{g}F^JN^{,i} +
2\sqrt{g}NM^{JK}A_K^i + \pounds_{\xi}\pi^{Ji}.
\end{array}
\ee

The evolution of the Hamiltonian constraint is then
\be
\begin{array}{c}
\frac{\partial}{\partial\lambda}\left(\sqrt{g}(R + U_{\mbox{\scriptsize A\normalsize}_I}) 
- \frac{1}{\sqrt{g}}\left(p^{ij}p_{ij} - \frac{1}{2}p^2 + \pi^I_i\pi_I^i\right)\right) = \\
2N_{,j}\left(2{p^{ij}}_{;i} - (\pi^{Ic}({A_{Ic}}^{;j} - {{A_I}^j}_{;c}) - {\pi_{Ic}}_{;c}A^{Ij})\right) +
N\left(2{p^{ij}}_{;i} - (\pi^{Ic}({A_{Ic}}^{;j} - {{A_I}^j}_{;c}) - {\pi_{Ic}}_{;c}A^{Ij})\right)_{;j} \\
+ \frac{Np}{2\sqrt{g}}\left(\sqrt{g}(R + U_{\mbox{\scriptsize A\normalsize}_I}) 
- \frac{1}{\sqrt{g}}\left(p^{ij}p_{ij} - \frac{1}{2}p^2 + \pi^I_i\pi_I^i \right)\right)
+ \pounds_{\xi}\left(\sqrt{g}(R + U_{\mbox{\scriptsize A\normalsize}_I}) 
- \frac{1}{\sqrt{g}}\left(p^{ij}p_{ij} - \frac{1}{2} p^2 + \pi^I_i\pi_I^i\right)\right)\\
+\frac{1}{N} \left((4C_1O^{IK} + \delta^{IK})(N^2\pi_I^a{A_{Ka}}^{;b})_{;b}
+ (4C_2O^{IK} - \delta^{IK})(N^2\pi_I^a{A_K^b}_{;a})_{;b}
+ 4C_3O^{IK}(N^2\pi_I^a{A_K^b}_{;b})_{;a} \right)\\
- \frac{1}{N} (N^2{\pi_K^a}_{;a}A^{Kb})_{;b} 
+ \frac{2}{N} \bar{C}^{abcd}{B^I}_{JK}  (N^2 \pi_{Ia}A^J_cA^K_d)_{;b}
+ \frac{2}{N} \epsilon^{abc}D_{IK}(N^2\pi^I_aA^K_c)_{;b} + \frac{2}{N}F^I(N^2\pi_{Ii})^{;i} \\
- \frac{2}{N} O^{IK}\left(N^2 \left( p_{ij} - \frac{p}{2} g_{ij} \right) A_{K(b;d)} 
(2A_I^iC^{ajbd} - A_I^aC^{ijbd}) \right)_{;a}  \\
- \frac{1}{N} {B^I}_{JK} \left( N^2 \left( p_{ij} - \frac{p}{2} g_{ij} \right) A^J_b A^K_d ( 2A_I^i\bar{C}^{ajbd} - A_I^a\bar{C}^{ijbd}) \right)_{;a}  \\
- \frac{1}{N} F^I \left( N^2 \left( p_{ij} - \frac{p}{2} g_{ij} \right) (2A_I^ig^{aj} - A_I^ag^{ij}) \right)_{;a}
\label{evolham}.
\end{array}
\ee 
We demand that this vanishes weakly.  The first four terms vanish weakly by definition, 
leaving us with ten extra terms.  
Suppose that these do not to automatically vanish; 
then we would require new constraints.  
We shall deal with this possibility 
by implementing our interpretation of Dirac that we presented in Sec. 2.
In this case we have at most $2 + 3K$ degrees of freedom, so if we had $3K$ or more new constraints, 
the vector field theory would be trivial. 
Furthermore, all constraints must be independent of N.  
Thus, terms in $N^{;a}$ must be of the form $(N^{;a}V_{Ja})S^J$ for the theory to be nontrivial 
(and we cannot have more than $3K$ independent scalar constraint factors $\{S\}$ in total).  
Most of these scalars will vanish strongly, which means they will fix coefficients in the potential ansatz. 
Finally, (\ref{evolham}) is such that all the non-automatically vanishing
terms in $N$ are partnered by terms in $N^{;a}$.  So the above big
restriction on the terms in $N^{;a}$ affects all the terms.  
The above argument applied to (\ref{evolham}) may be conveniently subdivided into the following three steps.

1) The first, second, third, sixth and seventh extra terms have no
nontrivial scalar factors, so we are forced to have $O^{IK}  =
\delta^{IK}$, $C_1 = -C_2 = -1/4$, $C_3 = 0$, $D_{IK} = 0$ and
$F_I = 0$.  The conditions on the $C$'s correspond to the
vector fields obeying the gravitational light-cone. 

2) This automatically implies that the eighth and tenth terms also vanish.   
The only nontrivial possibilities for the vanishing of the ninth term are
$\bar{C}_3 = 0$ and either $B_{I(JK)} = 0$ or 

\noindent $\bar{C_1} = -\bar{C_2} \equiv -\mbox{\sffamily g\normalfont}/4$, 
say. In fact, because of the symmetry properties of this ansatz term, the second and third of these properties 
imply each other if the first holds.  

3) This finally leaves us with $K$ new scalar constraint factors from the fourth and fifth terms, 
\be
{\cal G}_J \equiv {\pi_J^a}_{;a} - \mbox{\sffamily g\normalfont}
B_{IJK}\pi^I_aA^{Ka} \approx 0. 
\ee

Next, we examine the evolution of this internal-index vector of new constraints
\be
\begin{array}{c}
\frac{\partial}{\partial\lambda}({\pi_J^a}_{;a} - \mbox{\sffamily g\normalfont} B_{IJK}\pi^I_aA^{Ka}) = \\
\pounds_{\xi}({\pi_J^a}_{;a} - \mbox{\sffamily g\normalfont}
B_{IJK}\pi^I_aA^{Ka})
-\frac{2N}{\sqrt{g}} \mbox{\sffamily g\normalfont} \pi^K_i\pi^{Ii}B_{IJK} \\
+ \frac{\sqrt{g}}{2}\mbox{\sffamily g\normalfont} A^{Ki;b}(A_{Ii;b} - A_{Ib;i}){B^I}_{JK} \\
+ \sqrt{g}(NA^{dK}A^{iL}A^M_d)_{;i}\left( \begin{array}{l} \bar{\bar{C_1}}(I_{JKLM} + I_{KJML} + I_{LMJK} + I_{MLKJ})\\
+ \bar{\bar{C_2}}(I_{JKML} + I_{KJLM} + I_{MLJK} + I_{LMKJ}) \\
+ 2\bar{\bar{C_3}}(I_{(JL)KM} + I_{KM(JL)})
-  \frac{1}{2} \mbox{\sffamily g\normalfont}^2 {B^I}_{JK}B_{IML}  \end{array} \right)  \\
- \frac{\sqrt{g}}{2}  \mbox{\sffamily g\normalfont}^2  N A^{Li} A^{Mb} A^K_{b;i} ({B^I}_{JK}B_{ILM} + {B^I}_{JM}B_{IKL} +{B^I}_{JL}B_{IMK}) \\
- {B^Q}_{JP} \mbox{\sffamily g\normalfont} A^{iP}A^{Kd}A^L_iA^M_d
\left( \begin{array}{l} \bar{\bar{C_1}}(I_{QKLM} + I_{KLMQ} + I_{LKQM} + I_{LKMQ})  \\
+ \bar{\bar{C_2}}(I_{KQLM}  + I_{KLQM} + I_{QKML} + I_{KQML}) \\
+ 2\bar{\bar{C_3}}(I_{KM(QL)} + I_{(QL)KM} ) \end{array} \right)\\
+ 3\epsilon^{ibc}\left( E_{JNK}(NA^K_cA^N_b)_{;i} + \mbox{\sffamily g\normalfont} E_{QNK} N A^K_c A^N_b A^P_i {B^Q}_{JP} \right) \\
+ 2\sqrt{g}\left(M_{JK}(NA^{Ki})_{;i} - \mbox{\sffamily
g\normalfont} M_{QK} {B^Q}_{JP}NA^{Ki}A^P_i \right)

\label{propagnew}
\end{array}
\ee and we demand that this vanishes weakly.  The first term vanishes weakly by definition,
leaving us with seven extra terms.  
Again, we first consider the $N^{;a}$ parts of the terms, 
which lead us to the following extra steps.

4) For the theory to be nontrivial, the third, sixth and seventh 
non-automatically vanishing terms force
us to have, without loss of generality, 
$I_{JKLM} = {B^I}_{JK}B_{ILM}$, $\bar{\bar{C}}_2 = - \bar{\bar{C}}_1 =
\mbox{\sffamily g\normalfont}^2 /16$, $\bar{\bar{C}}_3 = 0$,
$E_{JNK} = 0$ and $M_{JK} = 0$.  
This last condition means that the interacting vector fields must be fundamentally massless.  

5) We are then left with the first, second, fourth and fifth terms.  
The fourth term forces upon us the Jacobi identity \be
{B^I}_{JK}B_{ILM} + {B^I}_{JM}B_{IKL} + {B^I}_{JL}B_{IMK} = 0. \ee
Thus, the $B_{IJK}$ are axiomatically the structure
constants of some Lie algebra, $\bf {\cal A} $ \normalfont.
Furthermore, the vanishing of the first term forces us to have
$B_{IJK} = B_{[I|J|K]}$, which means that the $B_{IJK}$ are totally antisymmetric.  
Finally, the second and fifth terms are then automatically zero. So the potential term
must be \be U_{\mbox{\scriptsize A\normalsize}_I} = -
\frac{1}{8}(A^I_{a;b} - A^I_{b;a} + \mbox{\sffamily
g\normalfont}{B^I}_{JK}A^J_aA^K_b) (A_I^{a;b} - A_I^{b;a} +
\mbox{\sffamily g\normalfont} B_{ILM}A^{La}A^{Mb}). \ee

We will now investigate what the total antisymmetry of $B_{IJK}$ means.  
In the standard approach to Yang--Mills theory in flat spacetime, 
one starts with Lorentz and parity invariance, which restricts the Lagrangian to be 
$\mbox{\sffamily L\normalfont}_{\mbox{\scriptsize A\normalsize}_I}^{(4)} =
-Q_{AB}F^A_{\mu\nu}F^{B\mu\nu}$, where \be F^A_{\mu\nu} =
(A^A_{\mu;\nu} - A^A_{\nu;\mu} + \mbox {\sffamily
g\normalfont}_{\mbox {\scriptsize
c\normalsize}}{f^A}_{JK}A^J_{\mu}A^K_{\nu}) \ee is the field
strength tensor, ${f^I}_{JK}$ \normalfont are structure constants
and $\mbox {\sffamily g\normalfont}_{\mbox{\scriptsize
c\normalsize}}$ is a coupling constant. Furthermore, one demands
gauge invariance, $\delta \mbox {\sffamily
L\normalfont}_{\mbox{\scriptsize A}_I}^{(4)} = 0$, under the gauge
transformation \be A^I_{\alpha} \longrightarrow A^I_{\alpha} +
\mbox{\sffamily g\normalfont}_{\mbox{\scriptsize c\normalsize}}
{f^I}_{JK}\Lambda^J A^K_{\alpha} \ee which is equivalent to \be
Q_{(A|B}{f^B}_{C|D)} = 0. \label{a} \ee For $Q_{AB}$
positive-definite, Gell-Mann and Glashow \cite{GMG, Weinberg} have
shown that this and the following two statements are equivalent:
\be \exists \mbox{ basis in which} B_{ABC} = B_{[ABC]} \label{b}
\ee \be {\cal A } \mbox{ is a direct sum of compact simple and
U(1) Lie subalgebras}. \label{c} \ee Also, (\ref{a})
$\Leftrightarrow$ (\ref{b}) $\Leftrightarrow \dot{{\cal G}}_J
\approx 0 $ in the usual flat spacetime canonical working.

In contrast, we have started with $3$-dimensional vector fields on
$3$-geometries, obtained ${\cal H}$ as an identity and demanded
that $\dot {\cal H} \approx 0$, which has forced us to have the
secondary constraints ${\cal G}_J$.  But once we have the ${\cal
G}_J$, we can use them to do much the same as above. $\dot{\cal
G}_J \approx 0  \Leftrightarrow$ (\ref{b}) $\Leftrightarrow$
(\ref{a}), so our scheme allows the usual restriction (\ref{c}) on
the type of Lie algebra. We can moreover take (\ref{a}) to be
equivalent to the gauge invariance of $U_{\mbox{\scriptsize
A\normalsize}_I}$ under \be A^I_a \longrightarrow A^I_a + \mbox
{\sffamily g\normalfont} {B^I}_{JK}\Lambda^J A^K_a. \ee Thus, if we
introduce $K$ auxiliary variables $\Phi^K$ such that variation
with respect to them encodes ${\cal G }_K$ then we should do so
according to the best matching corresponding to this gauge
symmetry.  This uniquely fixes the form of $T_{\mbox{\scriptsize
A\normalsize}_I}(A_{Ia}, \Phi_J)$ to be \be T_{\mbox{\scriptsize
A\normalsize}_I} = g^{ad}(\dot{A}^I_a - \pounds_{\xi}A^I_a -
\Phi^I_{,a} + \mbox{\sffamily g\normalfont}{B^I}_{JK}A^J_a\Phi^K)                    
(\dot{A}_{Id} - \pounds_{\xi}A_{Id} - \Phi_{I,d} + \mbox{\sffamily
g\normalfont}B_{ILM}A^L_d\Phi^M) \label{YMKE}. \ee Finally,
if we identify $\Phi^K$ with $A_0^K$, we arrive at Yang--Mills
theory \cite{YM, PS, Weinberg} for $A_{\alpha}^K = [A_0, A_a]$, with coupling
constant $\mbox{\sffamily g\normalfont}$ and gauge group ${\cal
A}$ (corresponding to the structure constants
$B_{IJK}$). So this work constitutes a unique
derivation, from $3$-dimensional principles alone, of Yang--Mills
theory minimally-coupled to general relativity.

This last step is not immediate.  Picking $Q_{AB} = \delta_{AB}$,
the $(3 + 1)$ decomposition of $\mbox{\sffamily
L\normalfont}_{\mbox{\scriptsize A\normalsize}_I}^{(4)}$ yields
\cite{Ashtekar} \be T_{\mbox{\scriptsize A\normalsize}_I} =
g^{ad}(\dot{A}^I_a - {A_0^I}_{,a} + \mbox{\sffamily
g\normalfont}_{\mbox{\scriptsize c\normalsize}}{f^I}_{JK}A^J_a
A_0^K - \xi^mF^I_{am}) (\dot{A}_{Id} - A_{0I,d} +
\mbox{\sffamily g\normalfont}_{\mbox{\scriptsize
c\normalsize}}f_{ILM}A^L_d A_0^M - \xi^nF_{Ibn}). \ee One
must then integrate by parts and discard $\xi^mA_m^I{ \cal G }_I$
to show that this is equivalent to (\ref{YMKE}).

\section{Discussion}

\indent This work shows that the `Relativity without Relativity'
formalism can accommodate many examples of physical theories. We
can immediately write down a gravity-coupled formalism with the
$SU(3)$ gauge group of the strong force, or with larger groups
such as $SU(5)$ or $O(10)$, used in grand unified theories.
However, the work does not restrict attention to a single simple
gauge group, since it also holds for the direct sum (\ref{c}). One
can then rescale the structure constants of each U(1) or compact
simple subalgebra separately, which is equivalent to each
subalgebra having a distinct coupling constant \cite{Weinberg}.
The simplest example of this is to have $B_{IJK}$ = 0, which
corresponds to K non-interacting copies of electromagnetism. Other
examples include the $SU(2){\times}U(1)$ electroweak theory and
the $SU(3){\times}SU(2){\times}U(1)$ Standard Model. We stress
that our formalism does not have the power to single out what the
gauge groups of nature are.

BF\'{O} showed that a scalar field, a $3$-vector field, and a
$3$-vector field coupled to scalar fields all obey the same
light-cone as gravity.  In this paper we have shown that this is
also true for $K$ interacting vector fields, so there is a
universal light-cone for bosonic fields, derived entirely from
$3$-dimensional principles.  Investigation of the fermionic sector
would tell us whether this light-cone is indeed universal for all
the known fields of nature.  We also note that our formalism
reveals that the universality of the light-cone and gauge theory
have a common origin resulting from the universal application of
best matching to implement 3-diffeomorphisms in conjunction with
the need to propagate the quadratic Hamiltonian constraint.

In our $3$-dimensional formalism, fundamental vector fields are
not allowed to have mass. The only bosonic fields allowed to have
mass are scalar fields. This would make spontaneous symmetry
breaking a necessity if we are to describe the real world, since
the $W^+$, $W^-$ and $Z$ bosons, believed to be responsible for
the weak force, are massive. Moreover, if the study of fermions in
our formalism were to reveal these to be also fundamentally
massless, one would have a $3$-dimensional derivation that mass
necessarily arises from Higgs scalar fields.

It would be interesting to consider whether our formalism can
accommodate topological terms. Although it is not free of
controversy \cite{Peccei, PS}, t'Hooft's standard explanation of
the low energy QCD spectrum makes use of an extra topological
term \be \frac{\Theta^2\mbox{\sffamily g\normalfont}
_{\mbox{\scriptsize strong \normalsize }}^2}{32{\pi}^2}
\epsilon_{\alpha\beta\gamma\delta}F^{\alpha\beta}F^{\gamma\delta}
\label{top} \ee in the classical Lagrangian \cite{Hooft}. The parameter $\Theta$
is constrained to be small by the non-observation of the neutron
dipole moment \cite{nedm}. The inclusion of this term corresponds to
dropping the parity-invariance of the Lagrangian. The new term is
a total derivative \cite{Weinberg}. Nevertheless it makes a
contribution to the action when the QCD vacuum is nontrivial.

Baierlein, Sharp and Wheeler discarded their gravitational total
derivatives. If kept, these would simply give a further additive
term: \be S_{\mbox{\scriptsize BSW \normalsize}} = \int d\lambda
\int d^3x\sqrt{g} \sqrt{ T_{\mbox{\scriptsize g\normalsize} }}
\sqrt{R} + \mbox{surface integral}. \ee One should thus treat
(\ref{top}) as another such additive term. So in our formalism,
just as in any other, we can, and should, allow for further
surface contributions to the action.  We argue also that the
accommodation of topological terms is not yet a problem for us,
because so far we are only describing a classical, unbroken,
fermion-free world. But the need for the new term arises from
quantum-mechanical considerations when quarks (which are
spin-$1/2$ fermions) are present. Furthermore, the quarks must be
massive in order for the four-dimensional theory to predict
$\Theta$-dependent physics \cite{Peccei}.

EA is supported by PPARC.  We would like to thank Brendan Foster, Domenico Giulini
and especially Niall \'{O} Murchadha for helpful discussions.

\end{document}